\documentclass[aps,prb,reprint,superscriptaddress]{revtex4-2}

\usepackage{graphicx}
\usepackage{color}
\usepackage{dcolumn}  
\usepackage{braket}
\usepackage{hyperref}

\begin{document}

\def\lsco{La$_{1-x}$Sr$_{x}$CoO$_3$}
\def\lco{LaCoO$_3$}
\def\etal{{\it et al.}}

%\preprint{}


\title{
	\boldmath
Photo-induced insulator-to-metal transition and coherent acoustic phonon propagation in LaCoO$_3$ thin films explored by femtosecond pump-probe ellipsometry\unboldmath}


\author{M. Zahradn\'\i k}
\affiliation{ELI Beamlines, Fyzik\'aln\'\i\ \'ustav AV \v{C}R, v.v.i., Za Radnic\'\i\ 835, 25241 Doln\'\i\ B\v{r}e\v{z}any, Czech Republic}

\author{M. Kiaba}
\affiliation{Department of Condensed Matter Physics, Faculty of Science, Masaryk University, Kotl\'a\v{r}sk\'a 2, 611 37 Brno, Czech Republic}

\author{S. Espinoza}
\affiliation{ELI Beamlines, Fyzik\'aln\'\i\ \'ustav AV \v{C}R, v.v.i., Za Radnic\'\i\ 835, 25241 Doln\'\i\ B\v{r}e\v{z}any, Czech Republic}

\author{M. Rebarz}
\affiliation{ELI Beamlines, Fyzik\'aln\'\i\ \'ustav AV \v{C}R, v.v.i., Za Radnic\'\i\ 835, 25241 Doln\'\i\ B\v{r}e\v{z}any, Czech Republic}

\author{J. Andreasson}
\affiliation{ELI Beamlines, Fyzik\'aln\'\i\ \'ustav AV \v{C}R, v.v.i., Za Radnic\'\i\ 835, 25241 Doln\'\i\ B\v{r}e\v{z}any, Czech Republic}

\author{O. Caha}
\affiliation{Department of Condensed Matter Physics, Faculty of Science, Masaryk University, Kotl\'a\v{r}sk\'a 2, 611 37 Brno, Czech Republic}

\author{F. Abadizaman}
\affiliation{Department of Condensed Matter Physics, Faculty of Science, Masaryk University, Kotl\'a\v{r}sk\'a 2, 611 37 Brno, Czech Republic}

\author{D. Munzar}
\affiliation{Department of Condensed Matter Physics, Faculty of Science, Masaryk University, Kotl\'a\v{r}sk\'a 2, 611 37 Brno, Czech Republic}

\author{A. Dubroka}
\email[]{dubroka@physics.muni.cz}
%\thanks{}

\affiliation{Department of Condensed Matter Physics, Faculty of Science, Masaryk University, Kotl\'a\v{r}sk\'a 2, 611 37 Brno, Czech Republic}


\date{\today}

\begin{abstract}
We have studied ultrafast dynamics of thin films of LaCoO$_3$ and La$_{0.5}$Sr$_{0.5}$CoO$_3$ with femtosecond pump-probe ellipsometry in the energy range of 1.6--3.4~eV. We have observed a large pump-induced transfer of spectral weight in LaCoO$_3$ that corresponds to an insulator-to-metal transition. The photo-induced metallic state initially relaxes via a fast process with a decay constant of about 200~fs. Both LaCoO$_3$ and La$_{0.5}$Sr$_{0.5}$CoO$_3$ exhibit a significant secondary transient structure in 
the 1--30~ps range.  Results of measurements on films with different thicknesses demonstrate that it corresponds to a propagation of an acoustic strain pulse. On timescales longer than 100~ps, heat diffusion to the substrate takes place that can be modelled with a bi-exponential decay.
\end{abstract}
\pacs{xxx}
%71.30.+h 	Metal-insulator transitions and other electronic transitions
%75.47.Lx 	Magnetic oxides
%75.50.Cc 	Other ferromagnetic metals and alloys
\maketitle

\section{Introduction}
Insulator-to-metal (IM) transitions are intriguing phenomena involving  huge resistivity changes of many orders of magnitude. The transition between the insulating and the metallic ground state is typically achieved by doping, change of temperature, pressure or chemical composition, or by magnetic field~\cite{Imada1998}. In \lco,  an IM transition can be induced both by doping and by changing temperature. With hole doping, usually achieved by the exchange of trivalent La$^{3+}$ by a divalent ion, e.g., Sr in the \lsco\ compound, a ferromagnetic metallic state is developed for $x>0.18$~\cite{Wu2003,SamalKumar}. 
\lco\ is a diamagnetic insulator below 50~K, however, in the intermediate temperature range between 100 and 400~K, it exhibits semiconducting and paramagnetic behavior, and above about 500~K it turns into a bad metal~\cite{Tokura1998}. The physics of the cobaltites is considerably complicated by a quasi-degeneracy between the low-spin (LS, $t_{2g}^6$ $e_g^0$), the intermediate spin (IS, $t_{2g}^5$ $e_g^1$) and the high-spin (HS,$t_{2g}^4$ $e_g^2$) state of a Co ion. This is due to a competition between the Hund's rule coupling and the crystal field splitting~\cite{Maekawa2004}. The question of which spin state dominates in  the cobaltites has been a subject of an intense debate lasting over many decades~\cite{Degroot1990,Korotin1996,Zobel2002,Ishikawa2004,Yan2004,Haverkort2006,Ropka2003,Podlesnyak2006,Merz2010,Krapek2012,Hariki2020}.

Femtosecond pump-probe spectroscopies~\cite{Sundaram2002} have been used to study the pump-induced IM transitions in many materials, e.g. in cuprates~\cite{Matsuda1994}, vanadates~\cite{Vikhnin2006}, nickelates~\cite{Torriss2018,Iwai2003} and  organics~\cite{Okamoto2007}. 
Concerning cobaltites, Okimoto \etal~\cite{Okimoto2009} found transient features on picosecond timescales in Pr$_{0.5}$Ca$_{0.5}$CoO$_3$ that were interpreted, based on model calculations, as a consequence of a propagation of a photonically created metallic domain at the velocity of the ultrasonic wave. Izquierdo~\etal\cite{Izquierdo20019} examined \lco\ using femtosecond soft x-ray spectroscopy and found picosecond transient features that were interpreted in terms of a model involving several steps of the bulk pump-induced metalization process. Opinions on which transient features are due to bulk processes and which are due to a propagation of a domain or wave remain contradictory. We reexamine the pump-induced optical response of cobaltites with femtosecond ellipsometry that was recently developed~\cite{Rebarz2017,Espinoza2019,Richter2020,Richter2021}.
Ellipsometry is a self-normalizing technique that allows one to determine very accurately and reproducibly the complex dielectric function without a need for Kramers-Kronig analysis. 
Here we report on the pump-probe ellipsometric study of \lsco\ thin films with $x=0$ and $x=0.5$.  In the $x=0$ compound, we have observed a large pump-induced redistribution of the optical spectral weight from high to low energies that is indicative of the IM transition. 
In order to discern which transient features correspond to bulk phenomena and which to a propagation of a wave, we have investigated several thin films with different thicknesses. The data show that the dynamics on the picosecond timescale probed using visible wavelengths is a consequence of a strain pulse (the so-called coherent acoustic phonon) propagation~\cite{Thomsen1986} possibly accompanied by an increased HS concentration.

\section{Experiment}
\label{Experiment}
Several films of LaCoO$_3$ and La$_{0.5}$Sr$_{0.5}$CoO$_3$ were grown by pulsed laser deposition on 10~$\times$~10~mm$^2$ substrates (La$_{0.7}$Sr$_{0.3}$) $\times$ (Al$_{0.65}$Ta$_{0.35}$)O$_3$ (LSAT) at 700$^\circ$~C and  0.1~mbar oxygen partial pressure using fluency of 2~J/cm$^2$ of  the excimer laser with wavelength 248~nm. The samples were post-annealed at 550~$^\circ$C for 3 hours in room pressure oxygen atmosphere to decrease oxygen-vacancy concentration. X-ray diffraction measurements confirmed that the films are epitaxial. The thickness of the films, $d$,  was determined using X-ray reflectometry and ellipsometry.

Time-resolved spectroscopic ellipsometry measurements were performed using femtosecond pump-probe ellipsometer at~ELI~Beamlines~\cite{Rebarz2017}. The system was based on~an amplified Ti:sapphire laser (Coherent Astrella) with its fundamental mode of~35~fs pulses at~800~nm with 1~kHz repetition rate and pulse energy of~6~mJ. 
About $10\ \mu$J of~the fundamental beam was employed as a pump beam and focused on a sample so that the fluency was about 3~mJ/cm$^2$. About $1\ \mu$J was used to~generate supercontinuum white-light in~a CaF$_2$ window, which served as a probe beam allowing to~investigate the spectral range from about 1.6 to about 3.4~eV.
The measurements were carried out at the angle of~incidence of $\alpha=60^{\circ}$ and  the angle of incidence of the pump beam was $55^{\circ}$.
In~\textit{Polarizer-Sample-Compensator-Analyzer} configuration with rotating compensator, transient reflectance-difference spectra were measured by scanning the pump-probe delay. The data were acquired repeatedly for multiple different azimuth angles of the compensator while the polarizer and analyzer were kept fixed at~$\pm45^{\circ}$. 
In~order to~calculate the ellipsometric angles from the series of~measurements at~different compensator angles, the M\"{u}ller matrix formalism was employed for each photon energy and delay time, where the obtained reflectance-difference spectra were processed by Moore-Penrose pseudo-inversion using reference equilibrium-state spectra. Further details on~the experimental setup, as well as the data evaluation procedure, can be found elsewhere~\cite{Espinoza2019,Richter2020,Richter2021}.
The equilibrium-state ellisometric data were measured using Woollam VASE ellipsometer in the 0.6 -- 6.5~eV range and Woollam IR-VASE ellipsometer in the 0.05 -- 0.6~eV range.

\begin{figure}[t]
	\centering
	\vspace*{-0.1cm}
	\hspace*{-0.9cm}
	\includegraphics[width=10cm]{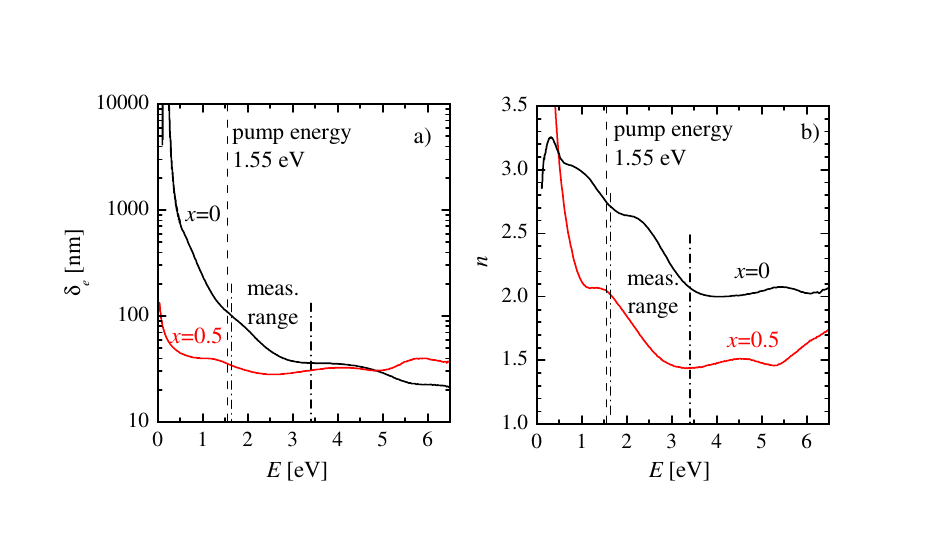}
	\vspace*{-0.2cm}
	\caption{Penetration depth (a) and the index of refraction (b) of \lsco\ thin films with $x=0$ (black solid line) and $x=0.5$ (red solid line). The dashed and dashed-dotted lines denote the pump energy and limits of the measurement range, respectively.}
	\label{PenDepth}
\end{figure}

All ellipsometric data were analyzed using the standard model of coherent interferences in a thin film on a substrate~\cite{HandbookElli}. The optical response of the substrate was measured on a bare substrate. The substrate did not exhibit any significant pump-induced response. 
The dielectric function employed in the analysis of the pump-probe ellipsometry data of thin films was modeled as a sum of the Lorentz oscillators~\cite{HandbookElli}. The equilibrium ellipsometric data were analyzed at each photon energy (the so-called point by point fit). 
Following this procedure, the obtained equilibrium spectra represent the optical response of a thin film that is in principle independent of the film thickness. In case of the pump-probe ellipsometry, some aspects of the data similarly reflect the change of the optical constants of the material and should be independent of the film thickness
for thicknesses sufficiently smaller than the pump penetration depth. However, additionally, the data may involve features corresponding, e.g., to propagation of a strain pulse~\cite{Thomsen1986,Ishioka2019} and to thermal diffusion~\cite{Sundaram2002}, that are in principle thickness dependent. Consequently, the obtained optical spectra represent pseudo-optical constants reflecting the overall response of the heterostructure. One possibility of how to discern one type of phenomena from the other is by examining films with various thicknesses as we do in the present work. Figures~\ref{PenDepth}(a) and \ref{PenDepth}(b) display the penetration depth, $\delta_e$,  and the real part of index of refraction, $n$, respectively, 
of \lsco\ thin films with $x=0$ and $x=0.5$ obtained using the equilibrium ellipsometry. Penetration depth was determined from the imaginary part of the index of refraction $k$ as $\delta_e=\lambda/(2\pi k)$, where $\lambda$ is the vacuum wavelength of the radiation. The penetration depth at the energy of the pump of 1.55~eV (dashed vertical line) is about 110 and 35 nm for $x=0$ and $x=0.5$, respectively. 

In the rest of the paper, we display the obtained (pseudo)-optical constants in terms of the real part of the optical conductivity, $\sigma_1$, related to the imaginary part of the dielectric function $\varepsilon_2$ as $\sigma_1(\omega)=\varepsilon_0\omega\varepsilon_2(\omega)$.  The quantity $\sigma_1$ is used to describe the absorption of radiation and it obeys the conductivity sum rule~\cite{Dressel}. The information represented by $\sigma_1$ is complemented by that contained in the real part of the dielectric function, $\varepsilon_1$, representing the so-called inductive response.

\begin{figure*}[t]
	\centering
	\vspace*{-0.1cm}
	\hspace*{-0.5cm}
	\includegraphics[width=14cm]{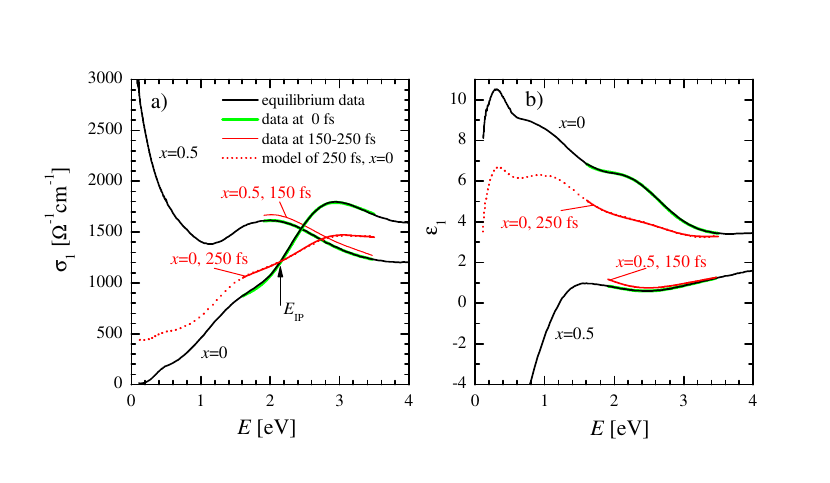}
	\vspace*{-0.2cm}
	\caption{
		Real part of the optical conductivity, $\sigma_1$ (a) and the real part of the dielectric function, $\varepsilon_1$ (b) of \lsco\ films with $x=0$, thickness 100~nm, and $x=0.5$, thickness 31~nm. Panels show the equilibrium data (black lines), the pump-probe data before the pump, i.e., at 0~fs delay (green thick solid lines) and at the delays corresponding to the largest pump-induced response of 250~fs ($x=0$) and 150~fs ($x=0.5$) (solid red lines). The red dotted lines represent Kramers-Kronig consistent extrapolations of the $x=0$ spectra at 250~fs to lower energies as detailed in the text. The arrow in panel (a)  denotes the so-called isosbestic point at $E=E_{\rm IP}$.}
	\label{Data}
\end{figure*}

\section{Data analysis and discussion}
Section~\ref{PeakResponse} is devoted to spectral analysis of the maximum pump-induced response which occurs near 200~fs. In the subsequent Sec.~\ref{TemporalEvolution}, we analyze the transient response in the whole range of delays, i.e., from 50~fs to 5~ns. 

\subsection{Maximum pump-induced response near 200~fs}
\label{PeakResponse}

Figure~\ref{Data}(a) displays the real part of the optical conductivity, $\sigma_1$, and Fig.~\ref{Data}(b) the real part of the dielectric function, $\varepsilon_1$,  of the equilibrium response of $x=0$ ($d=100$ nm) and $x=0.5$ ($d=31$ nm) thin films obtained by conventional ellipsometry (black solid lines). The pump-probe data of the two compounds before the pump (0~fs delay) are also displayed (green thick solid lines) and they essentially overlap with the equilibrium data. 
Similarly to previous reports~\cite{Tokura1998,Jeong-scientificreport},  the equilibrium spectrum of $\sigma_1$ of the parent compound ($x=0$)  exhibits a small band gap of about 0.2 eV and  pronounced interband transitions with a maximum near 3~eV. 
The conductivity spectrum of the doped sample ($x=0.5$) displays features similar to  those of $x=0.3$~\cite{Fris2018}, i.e., at low energies it exhibits a Drude peak describing the metallic response and at higher energies interband transitions with a band near 2~eV. The spectrum of $\varepsilon_1(x=0)$, see Fig.~\ref{Data}(b),  exhibits, below 1~eV, positive values in the range from 8 to 10 that are characteristic of an insulator, and the spectrum  of $\varepsilon_1(x=0.5)$  exhibits,  below 1.5 eV, a decrease to negative values with decreasing energy which is typical for a metallic response. 

Figures~\ref{Data}(a) and \ref{Data}(b) also display spectra at delays corresponding to maximum pump-probe response, i.e., at $250$~fs delay for the $x=0$ sample and at $150$~fs delay for the $x=0.5$ sample (red solid lines). The spectrum of $\sigma_1(x=0,250~\rm{fs})$ exhibits a significant suppression of the band at 3~eV and an increase of the conductivity below the crossing point (the so-called isosbestic point) at  $E_{\rm IP}=2.1$~eV. Clearly the pump causes a shift of optical spectral weight from energies above $E_{\rm IP}$ towards energies below. 
Recall that  the optical spectral weight per a frequency interval $(\omega_1, \omega_2)$ is defined as  $\int_{\omega_1}^{\omega_2}\sigma_1(\omega){\rm d}\omega$. 
The spectrum of $\varepsilon_1(x=0,250$~fs) is markedly decreased with respect to the equilibrium one [see Fig. \ref{Data}(b)], which, as we show below, also reflects the pump-induced shift of the spectral weight to lower energies. 

Further insight into the pump-induced response of the $x=0$ sample can be obtained by analyzing of the data with a Kramers-Kronig consistent model. 
We have modelled the equilibrium dielectric function as a sum of several Kramers-Kronig consistent contributions,
$\varepsilon(\omega)=\sum_j\varepsilon_{{\rm G},j}(\omega)+
\sum_j\varepsilon_{{\rm T-L},j}(\omega)$, where $\varepsilon_{\rm G}(\omega)$ stands for a Gaussian and $\varepsilon_{\rm T-L}(\omega)$ for a Tauc-Lorentz term~\cite{HandbookElli}. The fit of the model to the equilibrium data provides spectra of $\sigma_1$ and $\varepsilon_1$ essentially overlapping with those in Fig.\ref{Data} (fit not shown). For  the 250~fs data, the model dielectric function was complemented by the Drude term, $\varepsilon_{\rm D}(\omega)=-\omega_{\rm pl}^2/\omega(\omega+\rm i\gamma_D)$,  where $\omega_{\rm pl}$ is the plasma frequency and $\gamma_{\rm D}$ is the broadening parameter. 
We have fitted the model function to the data at 250~fs with only a limited number of free parameters (in particular, $\omega_{\rm pl}$, the parameters of the 3 eV band and of a band at higher energies) so that the overall shape of the response is conserved. The resulting model spectra (displayed as red dotted lines in Figs.~\ref{Data}(a) and \ref{Data}(b)) reveal that the conductivity and thus the spectral weight increases even below the lowest energy of our measurements of 1.6~eV. The fits are not sensitive to details of the low energy conductivity, such as the broadening of the Drude term $\gamma_{\rm D}$, which was fixed at $1$~eV. Nevertheless, it is reasonably sensitive to the increase of the low energy spectral weight, which is essentially given by $\omega_{\rm pl}^2$, for which the fitting yielded $\omega_{\rm pl}^2=(3.3\pm0.1$)~eV$^2$. 
This sensitivity comes predominantly from the pump-induced decrease of  $\varepsilon_1(x=0)$ [see Fig.~\ref{Data}(b)], which is a typical signature of a formation of a metallic (Drude) contribution. The corresponding number of charge carriers per cobalt ion can be calculated from $\omega_{\rm pl}^2$ using the standard Drude formula
$N=\epsilon_0m^*\omega_{\rm pl}^2a^3/e^2$, where $m^*$ stands for the effective mass and $a=3.8$~\AA\ is the lattice parameter. Provided $m^*$ equals the free electron mass, we obtain $N\approx0.13$, which is a sizable amount of charge per Co ion, comparable, e.g., to the one generated by photo-doping charge carriers in a halogen-bridged nickel-chain compound~\cite{Iwai2003}.

\begin{figure*}[t]
	\centering
	\vspace*{-0.1cm}
	\hspace*{-0.9cm}
	\includegraphics[width=19.7cm]{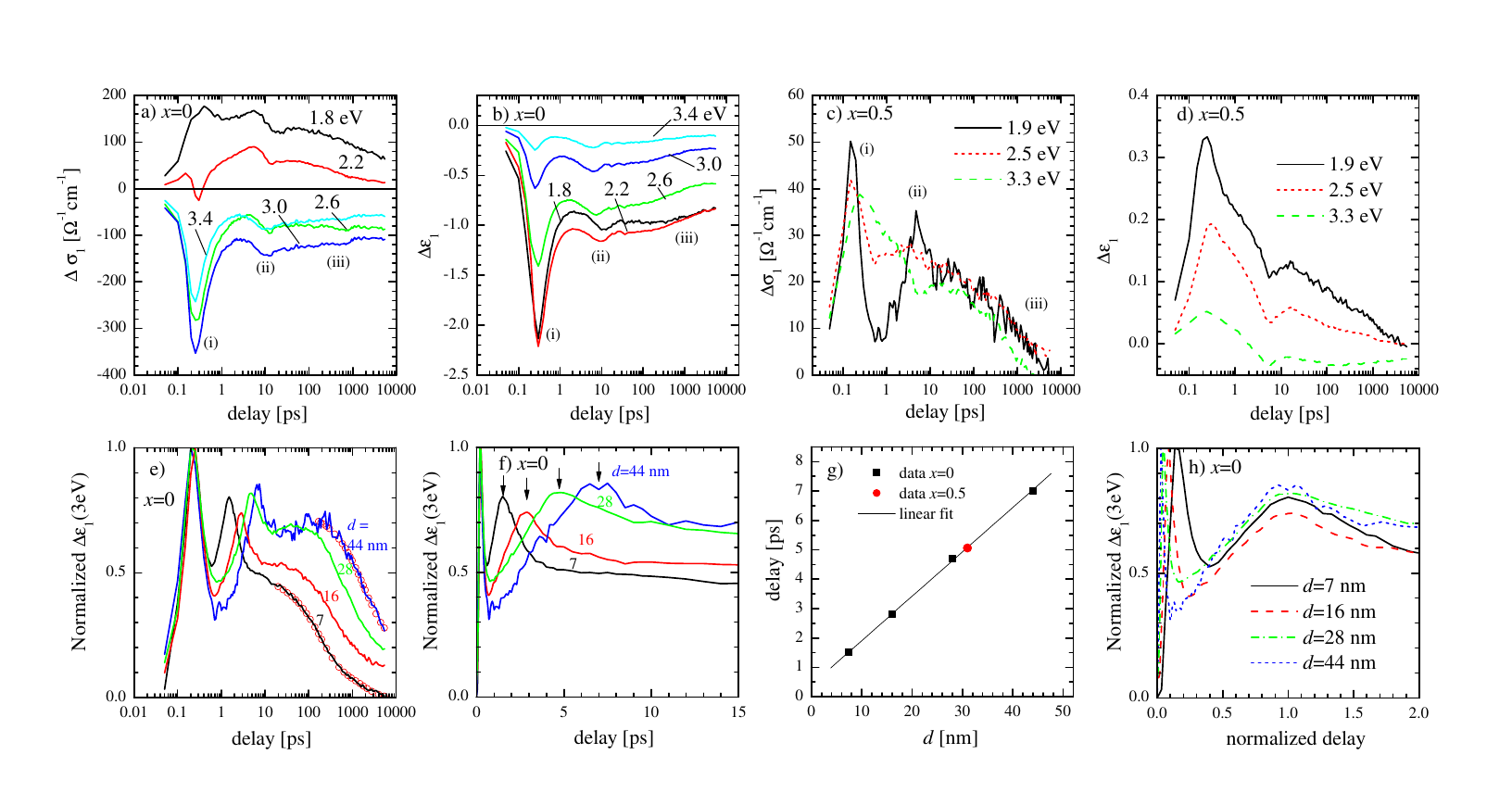}
	\vspace*{-0.2cm}
	\caption{
		Transient signals relative to the values before the pump of the real part of the conductivity, $\Delta\sigma_1$  [(a) and (c)] and of the real part of the dielectric function, $\Delta\varepsilon_1$ [(b) and (d)] of \lsco\ films with $x=0$, thickness 100~nm [(a) and (b)] and $x=0.5$, thickness 31~nm [(c) and (d)] at selected energies as shown in the graphs. Panels (e) and (f) show $\Delta\varepsilon_1$ transients at 3~eV normalized to the maximum values near 250~fs of several $x=0$ films with thicknesses ($d$) specified in the panels. Red circles in panel (e) represent fits by a bi-exponential decay model detailed in the text. Panel (g) displays the delays of the maxima denoted by arrows in panel (f) versus the film thickness $d$. Panel (h) displays transients from  panel (f) as a function of the delay normalized to that of the maximum.}	
	\label{Data3}
\end{figure*}

Regardless of the uncertainties about the details of the pump-induced spectrum of $\sigma_1$  below 1.6~eV (that call for an exploration in future), our analysis clearly shows the main trend: the pump induces a large transfer of spectral weight from energies above $E_{\rm IP}$ to lower energies, i.e., the material very likely becomes metallic. This trend can also be deduced simply from the measured pump-induced spectra in Fig.\ref{Data} 
[both $\sigma_1(x=0)$ and $\varepsilon_1(x=0)$].  They shift to some extent towards the spectra of the metallic $x=0.5$ sample. 
Such transfer of spectral weight can be expected because \lco\ exhibits a thermally induced IM  transition which is in the equilibrium optical response indeed accompanied by a transfer of spectral weight from high energies (above $E_{\rm IP}=1.4$~eV)  to lower energies giving rise to a Drude peak~\cite{Tokura1998}. In order to roughly estimate the pump-induced electronic temperature, we can compare our results with the equilibrium spectra obtained at high temperatures~\cite{Tokura1998}. The extrapolated DC conductivity at 250~fs of about 500~$\Omega^{-1}$cm$^{-1}$ corresponds to a temperature of the equilibrium spectra in the range between 500~K and 600~K~\cite{Tokura1998}. 

Concerning the $x=0.5$ sample, both $\sigma_1$(150~fs) and $\varepsilon_1$(150~fs) are slightly higher compared to the equilibrium spectra, see Fig.~\ref{Data}(a) and Fig.~\ref{Data}(b). A similar trend with increasing temperature in this energy range has been observed on metallic \lsco\ with $x=0.3$ (cf. Ref.~\cite{Fris2018}). Such changes at energies higher than the Drude isosbectic point of 0.3~eV~\cite{Fris2018} reflect an increase of the Drude scattering rate  $\gamma_{\rm D}$. The latter is expected for a metallic sample if charge carriers get comparably more disturbed by the excited quasiparticles. In other words, in the $x=0.5$ sample, the direction of the pump-induced spectral weight transfer is upward in energy and thus opposite to that of the $x=0$  sample.

\subsection{Analysis of the transients at time delays between 50~fs and 5~ns}
\label{TemporalEvolution}
Figures~\ref{Data3}(a) and~\ref{Data3}(b) display the transient optical constants relative to the values before the pump, $\Delta\sigma_1=\sigma_1(t)-\sigma_1(t=0)$ and $\Delta\varepsilon_1=\varepsilon_1(t)-\varepsilon_1(t=0)$, respectively, at selected energies for the $x=0$ film with $d=100$~nm. 
The transient signals exhibit the following general trend: (i) First they reach a peak value near 250~fs. This is followed by a fast relaxation with a characteristic decay time of about 200~fs. (ii) Another transient structure emerges between 1~ps and 30~ps forming a secondary maximum.
(iii) For times longer than 100~ps the signals monotonically decrease within the measurement delay span of 5000~ps. Note that the transients with the largest overall values, $\Delta\sigma_1$(3~eV) and $\Delta\varepsilon_1$(2.2~eV), have very similar profiles. This can be understood since, by virtue of the Kramers-Kronig relations, $\Delta\varepsilon_1$ at a given energy $E$ is a certain measure of the spectral weight transfered across $E$. Therefore, $\Delta\varepsilon_1$(2.2~eV) near $E_{\rm IP}\approx2.1$~eV follows the decrease of the 3~eV band [tracked by $\Delta\sigma_1$(3~eV)], whose spectral weight is partially transfered, after the pump, to energies lower than $E_{\rm IP}$. 

Figures~\ref{Data3}(c) and~\ref{Data3}(d) display the transient difference optical constants, $\Delta\sigma_1$ and $\Delta\varepsilon_1$, respectively, at selected energies, for the $x=0.5$  film with $d=31$~nm. Its $\Delta\sigma_1$(1.9~eV) transient, see Fig.~\ref{Data3}(c), exhibits (i) a peak near 150~fs that relaxes with decay time of about 100~fs  and also (ii) a pronounced secondary transient structure between 1~ps and 10~ps. Notably, these features are similar to the features (i) and (ii), of the thin film with $x=0$. The sign of both transients, however,  is here positive and the maximum values are almost an order of magnitude smaller than in the case of $x=0$.

The pump-probe response of a 100~nm \lco\ thin film was studied by Bielecki~\etal\ using  femtosecond reflectometry at 3.2~eV~\cite{Bielecki2014}. Their data are very similar to our $\Delta\varepsilon_1$ transients at 3.0 or 3.4~eV. They also observed the secondary transient structure (ii) and it was interpreted along the lines of a previous study by Okimoto \etal~\cite{Okimoto2009}. Authors of the latter study reported on a similar feature at picosecond timescale in the data of polycrystalline Pr$_{0.5}$Ca$_{0.5}$CoO$_3$ and concluded, on the basis of results of model computations, that the structure is due to a propagation of a metallic domain at a velocity of 4.4 km/s. This conclusion inspired us to test this hypothesis more directly by measuring the transient response on films with various thicknesses between 7 and 44~nm. The obtained transients $\Delta\varepsilon_1(x=0, 3$~eV), normalized to the value at the first peak near 250~fs, are shown in Fig.~\ref{Data3}(e) and Fig.~\ref{Data3}(f) on logarithmic and decadic scale, respectively. 
The data clearly show that the secondary transient structure (ii) occurs at a delay that increases with increasing film thickness. This is indeed a typical signature of a propagation of an acoustic strain  pulse between the surface and the interface in a thin film heterostructure, see e.g. Refs.\cite{Thomsen1984,Thomsen1986,Ishioka2019}. Figure~\ref{Data3}(g) demonstrates that the thickness dependence of the delay of the maximum [maxima marked by the arrows in Fig.~\ref{Data3}(f)] is a linear function.  Assuming, as in Ref.~\cite{Ishioka2019}, that the delay of the maximum corresponds to the arrival of the strain pulse to the interface, we obtain from each value a velocity and the average velocity amounts to $v=(5.7\pm0.5$)~nm/ps. Notably, the value of the $x=0.5$ film, obtained from the transient $\Delta\sigma_1$(1.9~eV) [see Fig.~\ref{Data3}(c)], represented in Fig.~\ref{Data3}(g) by the red circle, fits very well to the linear dependence, demonstrating that the optical signatures of the propagation of the strain pulse are universal in cobaltites, independent of whether the electronic ground state is metallic or insulating.

Interestingly, the width of the secondary peak of Fig.~\ref{Data3}(f) also scales with the film thickness. Figure~\ref{Data3}(h) displays these data as functions of the delay normalized to that of the maximum. All the transients here follow a similar triangular shape where the signal increases linearly until it reaches the maximum and then it gradually decreases. The triangular shape of the center of the detected dielectric response of the strain pulse was indeed predicted by Thomsen~\etal~\cite{Thomsen1986}. Note that cobaltites are very susceptible to strain; for example, the tensile  strain in LaCoO$_3$ thin films can drive them ferromagnetic~\cite{Fuchs2007}. The latter effect was explained using an effective attraction and stabilization of HS states in the tensile strained LaCoO$_3$ thin films~\cite{Sotnikov2020}. It is thus possible that in regions of positive strain in the pulse profile~\cite{Thomsen1986}, propagating in the normal direction, create a local anisotropic strain and an increase of the HS concentration occurs. This would manifest itself, in the case of LaCoO$_3$, in a shift of spectral weight to lower energies (corresponding to an increase of metallicity),  such as we observe in Figs.~\ref{Data3}(a) and \ref{Data3}(b). In the case of the metallic $x=0.5$ sample, provided that the same physical picture applies, the increased concentration of HS states would enhance scattering of the charge carriers and thus would decrease the metallicity, as we observe in Fig.~\ref{Data3}(c).

\begin{figure}[t]
	\centering
	\vspace*{-0.1cm}
	\hspace*{-0.9cm}
	\includegraphics[width=10.5cm]{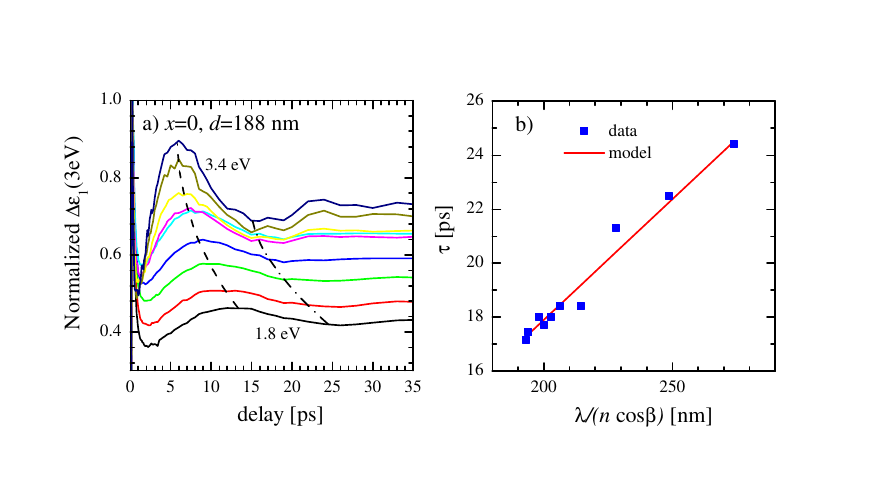}
	\vspace*{-0.2cm}
	\caption{a) Normalized $\Delta\varepsilon_1$ transients of a \lco\ film with thickness of 188~nm at selected energies  between 1.8 and 3.4~eV (with 0.2~eV steps). The dashed and dashed-dotted lines are guides to the eye. b) Period $\tau$ as a function of $\lambda/(n\cos\beta)$. The line corresponds to Eq.~(\ref{tau}) with $v=5.6$~nm/ps. }	
	\label{L73}
\end{figure}

The velocity of a strain pulse can also be determined from the period  of oscillations of the  transient signal in a film with large thickness (or in a bulk)~\cite{Thomsen1986,Fiebig2000,Ishioka2019}. These oscillations arise due to  interference involving the probe beam reflected from the front of the strain pulse. The profile has been predicted to be an oscillatory function with period~\cite{Grahn1989} 
\begin{equation}
\label{tau}
\tau=\frac{1}{2v}\frac{\lambda}{n\cos\beta}\;,
\end{equation}
where $n$ is the real part of the index of refraction and $\beta$ is the angle of refraction inside the film determined from the Snell's law as  $\cos\beta=\sqrt{1-\sin^2(\alpha)/n^2}$. Equation (\ref{tau}) was obtained using the expression for 
the phase factor occurring in the description  of interference in a thin film~\cite{HandbookElli}. In addition, the oscillatory profile is exponentially damped on the time scale given by $\delta_e/v$. 
Based on the above obtained velocity $v$, the expected $\tau$ for \lco\ amounts to 18--20~ps.
The samples of Fig.~\ref{Data3}(e) have small thicknesses so that the strain pulse reaches the interface before it gives rise to the oscillatory behavior. In order to explore the oscillatory transients in \lco, we have investigated a thicker film with $d=188$~nm. For such a relatively large thickness, the strain pulse reaches the interface at delays larger than 30~ps. Figure~\ref{L73}(a) displays the obtained normalized $\Delta\varepsilon_1$ transients at selected energies. The transients display the main secondary peak between 6~ps and 12~ps whose delay decreases with increasing probe energy (see the dashed line). The maximum is followed by a shallow minimum (see the dashed-dotted line). The oscillatory behavior is quickly damped because of the small penetration depth $\delta_e$ [see Fig.~\ref{PenDepth}(a)]. From the delays of this maximum and of the minimum, we can calculate the period of the oscillations $\tau$ and, using Eq.(\ref{tau}), the velocity $v$. The obtained average value of the velocity amounts to $v=(5.6\pm0.2$)~nm/ps which agrees very well with the value found above and with the value 5.9~nm/ps reported in Ref.~\cite{Bielecki2014}. The experimental values of $\tau$ (squares) are plotted in Fig.~\ref{L73}(b) as a function of  $\lambda/(n\cos\beta)$ together with the dependence given by Eq.~(\ref{tau}) with $v=5.6$~nm/ps. It can be seen  that that the relation (\ref{tau}) tracks the data very well.  
The transients displayed in Fig.~\ref{L73}(a) exhibit the trend observed earlier on picosecond timescales in Pr$_{0.5}$Ca$_{0.5}$CoO$_3$~\cite{Okimoto2009}: the maximum shifts to shorter delays with increasing energy of the probe. Our data show that this trend is not limited to the case of the first order insulator-metal transition of the latter compound but it is  common to cobaltites, including our metallic $x=0.5$ sample, and as we showed above, it corresponds to the propagation of a strain pulse possibly accompanied by an increased HS concentration. 

Next we address in more detail the metalization near 200~fs and the following fast relaxation (i). Figure~\ref{Data3}(e) demonstrates that, unlike the processes for delays longer than 1~ps, this feature is essentially independent of the film thickness and thus 
reflecting a bulk phenomenon occurring essentially homogeneously within the film. Note that the thicknesses of the films of Fig.~\ref{Data3}(e) are smaller that the pump penetration depth of about 110~nm. This fast relaxation can be modelled with an exponential decay (not shown), with the decay constant of about 200~fs. This is a typical relaxation time of excited electrons loosing their energy by emitting phonons~\cite{Sundaram2002}.  Near 1~ps, this fast relaxation is essentially over and the amount of spectral weight transferred across $E_{\rm IP}$ back to higher energies, as represented by $\Delta\varepsilon_1$(2.2~eV), in the 100~nm thick film [see Fig.~\ref{Data3}(b)], equals approximately 50\% of the maximum value. It is likely that at this delay, electron and phonon degrees of freedom of the film are essentially in thermal equilibrium~\cite{Sundaram2002}. The consecutive evolution of $\Delta\varepsilon_1$(2.2~eV) is only weakly modulated by the strain pulse propagation till about 100~ps when the film starts to cool down significantly as the heat diffuses to the substrate. 

The fast relaxation (i) in a \lco\ thin film has been also examined by Bielecki~\etal~\cite{Bielecki2014}. Similarly as in the present work, they found a fast decay on the 100~fs timescale and in addition its significant temperature dependence. Their analysis has shown that the data follow the Fermi statistics with the spin gap of 17~meV that seems to correspond to the HS--LS splitting~\cite{Hariki2020}. In a recent femtosecond soft x-ray spectroscopy study of bulk \lco, a fast relaxation with 170~fs decay time and a pronounced secondary maximum near 1.5~ps were reported~\cite{Izquierdo20019}. These features were interpreted as bulk phenomena due to several steps in the metalization process. The fast relaxation with the decay constant of 170~fs compares very well to the profile of our first peak (i). Concerning the maximum near 1.5~ps, the transients of our thickest films do not seem to exhibit any sharp structure near this delay, see Fig.~\ref{L73}(a) and Fig.~\ref{Data3}(b). Our data for delays larger then 1~ps, however, depend on the probe wavelength, therefore, it is not possible to compare them in a simple way with the data at soft x-ray wavelengths. 

The process (iii) on times scales larger than $100$~ps presumably corresponds to heat diffusion to colder parts of the sample~\cite{Sundaram2002}, most likely to the substrate which is transparent for the pump beam. The transients shown in Fig.~\ref{Data3}(e) demonstrate that the cooling process is slower for thicker films.
The reason is,  that with increasing film thickness, the total energy absorbed per surface area increases and the heat has a longer diffusion path before it reaches the substrate. In this delay range the transients can be well modelled by a bi-exponential decay, $A\exp(-t/t_A)+B\exp(-t/t_B)$.  This is documented by fits of the data for the $d=7$~nm and $d=44$~nm samples, represented by red circles in Fig.~\ref{Data3}(e). The obtained values of the relaxation times are $t_A=180$~ps and $t_B=1700$~ps for the 7~nm sample and $t_A=1800$~ps and $t_B=50$~ns for the 44~nm sample. For thicker films with $d=100$  and $188$~nm, the cooling process takes much longer than the upper delay of our measurement of 5~ns. A bi-exponential heat diffusion was also found in manganite films~\cite{Bielecki2010}.

\section{Summary}
The main findings of our investigations of the pump-probe optical response of \lco\ thin films can be summarized as follows: (i)  Within the first 200~fs after the pump, a transfer of optical spectral weight occurs, from high energies above 2.1~eV to lower energies, that we interpret as a consequence of the insulator-to-metal transition. 
The magnitude of the effect corresponds to about 0.13 electrons per Co ion.
Within the next about 200~ps, this photo-induced metallic state relaxes back whereby the conductivity differences decrease to about 50~\% of the maximum values. These features are essentially independent of film thickness and thus they correspond to a bulk excitation/relaxation within the penetration depth of the pump. (ii)  Between 1~ps and 30~ps additional features appear that are due to the propagation of an acoustic strain pulse at velocity of (5.6$\pm$0.2)~nm/ps. We observe similar phenomena also in the metallic La$_{0.5}$Sr$_{0.5}$CoO$_3$. Our data are compatible with the picture that the strain pulse is accompanied by an increased high-spin concentration. (iii) For delay times longer than 100~ps, we observe a heat film-to-substrate diffusion that can be modelled with a bi-exponential decay.

%\appendix
%\section{Acknowledgements}
\begin{acknowledgments}
We acknowledge fruitful discussions with J. Bielecki, F. F. Delatowski, D. Geffroy, and J. Kune\v{s}. We thank M. Golian for supporting measurements. This work was financially supported by the MEYS of the Czech Republic under the project  CEITEC 2020 (LQ1601) and by the Czech Science Foundation (GA\v{C}R)
under Project No. GA20-10377S. CzechNanoLab project LM2018110 funded by MEYS CR is gratefully acknowledged for the financial support of the measurements/sample fabrication at CEITEC Nano Research Infrastructure. ELI Beamlines authors acknowledge the support of the ELIBIO (No. CZ.02.1.01/0.0/0.0/15-003/0000447) and ADONIS (No. CZ.02.1.01/0.0/0.0/16-019/0000789) projects from the European Regional Development Fund. 
We acknowledge LM2017094, MEYS – Large research infrastructure project ELI Beamlines in Doln\'{i} B\v{r}e\v{z}any, Czech Republic, for provision of laser beamtime and would like to thank the instrument group and facility staff for their assistance. 
\end{acknowledgments}

\bibliographystyle{apsrev4-2}
\bibliography{bibliography}
%\nocite{Supplementary}
%\nocite{Okimoto1997}
%\nocite{PrietoRuiz2015}
\end{document}